\documentclass{article}

\usepackage{graphicx}
\usepackage{amsmath}

\newcommand\Rey{\mbox{\textit{Re}}}  

\title{Kinematics of Clustering}

\author{Steven Wang$^{1,3}$, 
Robert L. Stewart$^1$, 
Guy Metcalfe$^1$\thanks{Email address for correspondence: guy.metcalfe@csiro.au} and Jie Wu$^2$ \\
Commonwealth Scientific \& Industrial Research Organization (CSIRO)\\ 
Box 56, Highett VIC 3190 Australia\\
  $^1$Manufacturing Flagship\\
  $^2$Mineral Resources Flagship\\
$^3$Monash University, Department of Chemical Engineering\\ Box 36,
Clayton VIC 3800, Australia}

\date{8 July 2014}

\begin{document}

\maketitle

\begin{abstract}
  The dynamical system for inertial particles in fluid flow has both
  attracting and repelling regions, the interplay of which can
  localize particles.  In laminar flow experiments we find that
  particles, initially moving throughout the fluid domain, can undergo
  an instability and cluster into subdomains of the fluid when the
  flow Reynolds number exceeds a critical value that depends on
  particle and fluid inertia.  We derive an expression for the
  instability boundary and for a universal curve that describes the
  clustering rate for all particles.  
\end{abstract}


\section{Introduction}
\label{intro}

Dynamical systems theory is the natural language of transport.  The
kinematic equation $d\mathbf{x}/dt = \mathbf{u}(\mathbf{x},t)$
describes the Lagrangian motion of a passive particle at position
$\mathbf{x}$ moving according to fluid velocity field $\mathbf{u}$,
and for incompressible flows ($\nabla \cdot \mathbf{u} = 0$) this is a
volume-preserving dynamical system.  Remarkably the physical
coordinates of the fluid particles are exactly the same as the phase
space coordinates of the dynamical system, permitting direct
visualization of the system orbits, which can be regular or chaotic
\cite{Aref_prize_2002,sturman2006mathematical,Metcalfe_notes_2010}.
Particles, moving in fluids are often assumed to move passively if
their inertia is small enough, tending toward neutrally buoyant and
infinitesimally small.  A particle moves passively when it does not
change the fluid velocity field and instantaneously matches its own
velocity to that of the fluid.  How does the dynamical system for
particle motion change in the presence of finite inertia?

Here we visualize the motion of inertial particles using a stirred
tank laminar flow and examine a mechanism for particle localization
that arises from interaction of the intrinsic inertia of particle and
fluid.  Stirred tanks are used globally in the processing industries
and have been in industrial use almost unchanged for literally
centuries \cite{Agricola_DeReMetallica_1556}.  The new result is the
discovery and explanation of a particle clustering instability whereby
particles moving initially through the entire fluid volume
spontaneously move into and stay in a subdomain of the fluid.  The
explanation hinges on the existence and interplay of both attractors
and repellors in the augmented dynamical system for motion of inertial
particles.    As attractors and repellors are
caused by common features of natural or engineered flows, we
anticipate that this localization mechanism may be activated in many
laminar flows.

\section{Inertial Particle Dynamical System}
\label{sec:system}

\subsection{Fluid Motion}

Unbaffled stirred tank flow is axisymmetric in time-average, and so
produces a dynamical system that is 2-dimensional and conserves phase
space.  Figure~\ref{fig:skeleton}a shows the skeleton of the laminar
flow which has Kolmogorov--Arnold--Mosur (KAM) tubes above and below
the impeller.  The impeller provides a high frequency periodic
perturbation to the flow skeleton to create a sea of chaotic orbits
surrounding smaller KAM tubes (see figure~\ref{fig:results}).  The
boundaries of the KAM tubes are material surfaces, and fluid does not
cross them, except by slow diffusion.  This creates separated flow
regions that do not exchange fluid.  As stirred tanks are typically
driven to turbulence, only recently have studies of the chaotic
laminar flow in stirred tanks been reported
\cite{Fountain_3d_2000,Takahashi_object_2009,Wang_imr_2013}.

\subsection{Particle Motion}

Newton's law describing the velocity $\mathbf{V}_p(t)$ of a spherical
particle with density $\rho_p$ and radius $a$ immersed in a fluid with
density $\rho_f$ gives the Maxey--Riley (MR) equation of motion
\cite{1983_Maxey_equation}:
\begin{eqnarray}
\label{eq:MR}
\rho_p \frac{d\mathbf{V}_p}{dt} &=& \rho_f \frac{D\mathbf{u}}{Dt} + (\rho_p - \rho_f)\mathbf{g}
    - \frac{9 \nu \rho_f}{2 a^2} \left(\mathbf{V}_p - \mathbf{u} \right) \\ \nonumber
& & - \frac{\rho_f}{2}\left( \frac{d\mathbf{V}_p}{dt} - \frac{D\mathbf{u}}{Dt} \right).
\end{eqnarray}
The forces are, respectively, the force exerted by the undisturbed
flow on the particle, buoyancy force, Stokes drag and the added-mass
from part of the fluid moving with the particle; neglected are the
history force and higher order corrections
\cite{2012_Metcalfe_beyond}.  There is a subtle difference between
the operators $D/Dt$ and $d/dt$: $D/Dt = \partial /\partial t +
\mathbf{u} \cdot \mathbf{\nabla}$ is the usual convective derivative
taken along the path of a {\em fluid} element, while $d/dt = \partial
/\partial t + \mathbf{V}_p \cdot \mathbf{\nabla}$ is a derivative
taken along the {\em particle} trajectory.  Both derivatives are the
same for passive particles.  Inertia is characterized by the parameter
$\mu$,
\begin{equation}
\label{eq:inertia}
\mu = \frac{St}{R} \quad \mbox{with} \quad R = \frac{2 \rho_f}{ \rho_f + 2 \rho_p} \quad
\mbox{and} \quad St = \frac{2}{9} \left(\frac{a}{L}\right)^2 Re,
\end{equation}
where $St$ is the Stokes number, $R$ gives the density variation and
$\Rey$ is the flow Reynolds number; $\mu$ gives the ratio of the
particle relaxation time and the typical time scale of the
flow. Deviation of a particle from passive behavior is quantified by
the particle Reynolds number $\Rey_p = a |\mathbf{V}_p -
\mathbf{u}|/\nu$, $\nu$ the kinematic viscosity.  $\Rey_p = 0$ when a
particle moves passively with the fluid and becomes non-zero when
inertia causes particles to deviate from fluid streamlines: $\Rey_p
\rightarrow 0$ as $\mu \rightarrow 0$.

Using (\ref{eq:inertia}f) to non-dimensionalize (\ref{eq:MR}) gives
\begin{equation}
\label{eq:MR_reduced}
\mu \left(\frac{d\mathbf{V}_p}{dt} - \frac{3R}{2}\frac{D\mathbf{u}}{Dt} \right) = -\left(\mathbf{V}_p - \mathbf{u}\right),
\end{equation}
where we have neglected buoyancy by setting gravity to zero; although,
density differences are retained in $\mu$ through the added mass.
Equation~\ref{eq:MR_reduced} is valid in the limit of dilute particle
number density, $\Rey_p \ll 1$ and the assumption that particle motion
does not effect $\mathbf{u}$, i.e.\/ a one-way coupling.  As our
objective is the minimal model required to capture particle
clustering, agreement with data must ultimately approve the
approximations.


\begin{figure}
\centering
\begin{tabular}{cc}
\raisebox{-0.5\height}{\includegraphics[width=0.36\columnwidth]{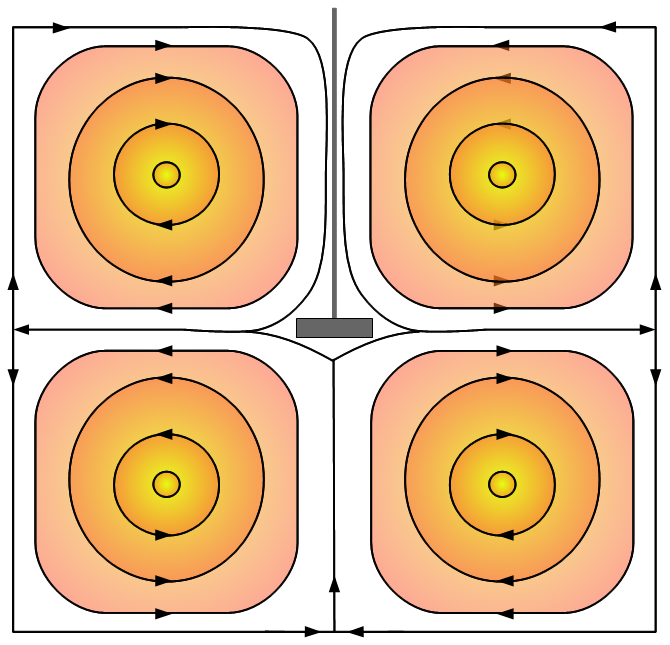}} &
\raisebox{-0.5\height}{\includegraphics[width=0.48\columnwidth]{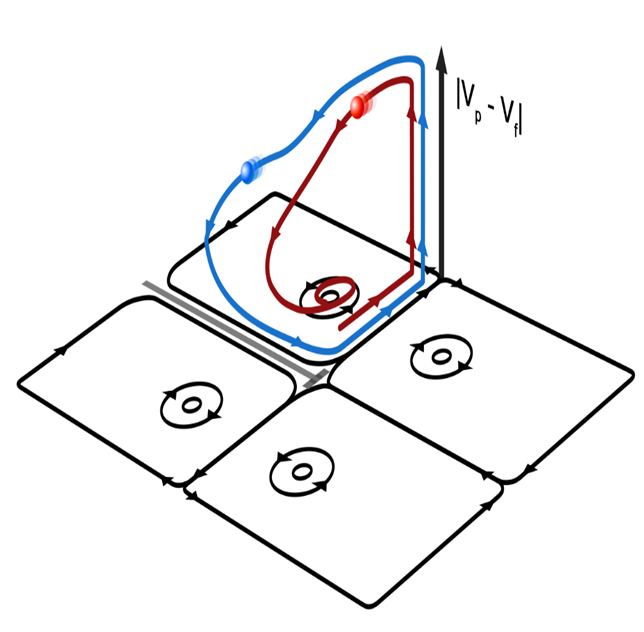}} \\
(a) & (b)
\end{tabular}
\caption{(a) Passive (fluid) particle dynamical system for stirred
  tank at low $\Rey$; two toroidal rolls exist above and below the
  impeller.  (b) Phase space of augmented dynamical system for
  particle motion.  Attracting hyperplane of fluid coordinates is
  perspective view of (a).  Vertical coordinate is $\Rey_p$.  Particle
  orbits repelled from the fluid hyperplane move in the full phase
  space and eventually reattract onto the hyperplane either outside
  tubes (blue) or inside tubes (red).}
\label{fig:skeleton}
\end{figure}

The addition of inertia augments the 2-dimensional fluid dynamical
system to 4 dimensions in which the inertial particle trajectories
move.  The dynamical system for motion in the particle phase space 
$\mathbf{\chi} = (\mathbf{X}_p, \mathbf{V}_p)$ is
\begin{subequations}  
\label{eq:inertial_system}
\begin{equation}
\frac{d\mathbf{X}_p}{dt} = \mathbf{V}_p
\end{equation}
\begin{equation}
\frac{d\mathbf{V}_p}{dt} =  -\mu^{-1} \left(\mathbf{V}_p - \mathbf{u}\right) 
                            + \frac{3R}{2}\frac{D\mathbf{u}}{Dt}.
\end{equation}
\end{subequations}
The inertial phase space has two key properties.  (1) There is an
attracting hyperplane containing passive fluid motion where
$\mathbf{V}_p = \mathbf{u}$.  This plane is overall attracting due to
fluid drag, but in places this plane is repelling.  Compact repelling
regions are called repellors in contradistinction to attractors.  (2)
The system (\ref{eq:inertial_system}) is non-conservative with a
divergence, the rate of contraction of phase space volume, given by
\begin{equation}
\label{eq:divergence}
\nabla \cdot  \mathbf{\chi} = -\mu^{-1}.
\end{equation}
Non-conservation of phase volume is generic when other forces are
added to the conservative fluid system \cite{2012_Metcalfe_beyond}.

Figure~\ref{fig:skeleton}b schematically shows the phase space and
representative orbits of the inertial dynamical system.  The oblique
plane is the fluid hyperplane and is the same as in
figure~\ref{fig:skeleton}a.  Due to hydrodynamic drag, a particle will
always eventually move nearly passively in the fluid hyperplane until
perturbed away.  When repelled from the fluid plane, a particle moves
non-passively through the rest of the augmented dynamical system.
This simply means that when a particle moves non-passively, it
requires four numbers, the particle location and its velocity, to
describe its state.  Because the material surfaces of the separated
flow regions are transport barriers {\em only} in the fluid plane,
when particles leave the fluid plane and move through particle
coordinate space (schematically shown in figure~\ref{fig:skeleton}b
with red and blue orbits), they can reattract to any part of the the
fluid plane, ending up inside the separated flow region (red orbit) or
outside of it (blue orbit).  When attracted into separated flow
regions that have no repellors, particles will stay in the attracting
subdomains.  This interplay between attracting and repelling flow
regions can lead to particle clustering, as we demonstrate below.

\section{Experiment}
\label{sec:experiment}


\begin{figure}
\centering
\includegraphics[width=\columnwidth]{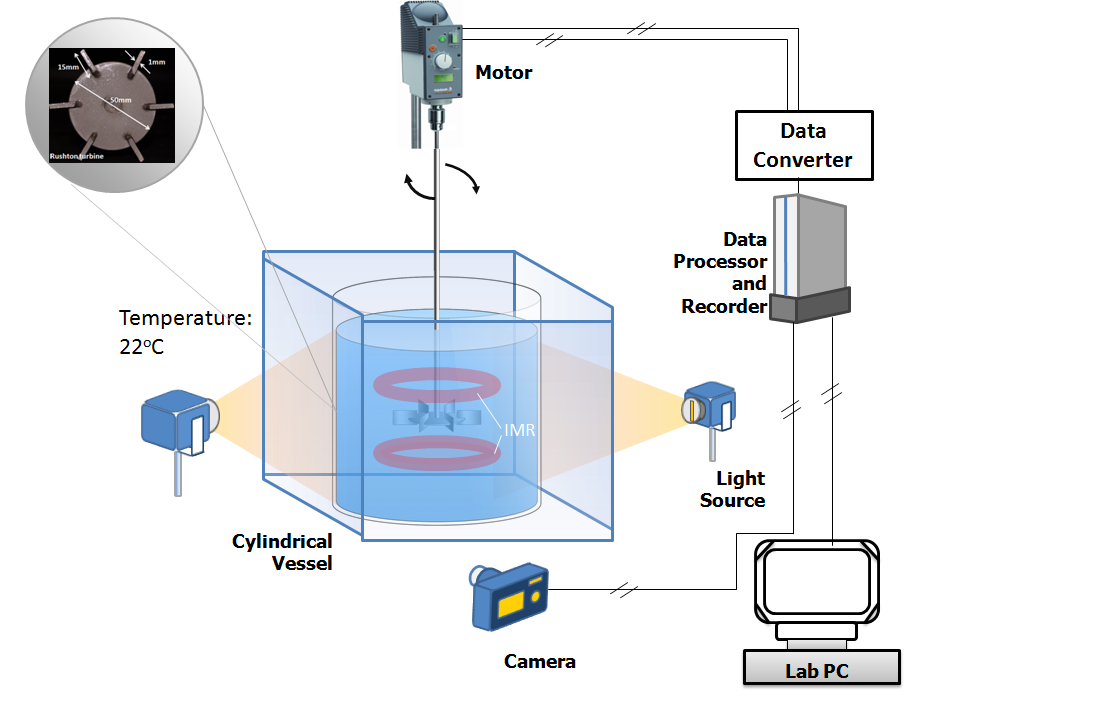}
\caption{A tank, filled with moderately viscous fluid and particles,
  is stirred at a controlled rate by a Rushton impeller (view along
  shaft, top left corner).  Light shines through a slit to illuminate
  a cross-section of flow which is filmed.}
\label{fig:rig}
\end{figure}

Our experiment (figure~\ref{fig:rig}) consists of an open tank (190~mm
diameter) filled (190~mm depth) with a moderately viscous fluid (1\%
water/glycerin; viscosity $\eta = 0.83$~Pa-sec; density $\rho_f =
1260$~kg/m$^3$).  Suspended by a centered shaft, half-way in the fluid
depth, is an impeller (6-bladed Rushton) of a type that sucks
fluid along the axial (shaft) directions and pushes fluid radially
outward in the impeller plane.  The impeller has a diameter $L =
70$~mm and is rotated at a controlled rate $\Omega$; the flow Reynolds
number is $\Rey = UL/\nu = \Omega L^2/\nu$, where $U = \Omega L$ and
$L$ are the characteristic velocity and length scales of the fluid
flow and $\nu = \eta/\rho_f$ is the kinematic viscosity.  $\Rey < 150$
here, an order of magnitude below the transition to turbulence.

\subsection{Particles}

\begin{table}
\begin{center}
\begin{tabular}{lccccc}
particle & $a$ & $\rho_p$ & $\rho_p/\rho_f$ & $a/L$ & $\mu/\Rey$ \\
         & [mm] & kg/m$^3$ & --- & $\times 10^2$ & $\times 10^4$ \\
\noalign{\smallskip}\hline\noalign{\smallskip}
Resin       & $0.65 \pm 0.05$ & 1220 & 0.97 & 0.93  & 0.28 \\
Polystyrene & $1.4 \pm 0.4$ & 1000 & 0.79 & 2.0 & 1.2 \\
Polystyrene & $2.3 \pm 0.4$ & 1000 & 0.79 & 3.3 & 3.1 \\
PMMA        & $3.18 \pm 0.05 $ & 1180 & 0.94 & 4.5 & 6.6 \\
Capsule     & $28 \pm 0.05$ & 874 & 0.69 & 40 & 430
\end{tabular}
\caption{Particle properties.  Fluid density $\rho_f = 1260$~kg/m$^3$
  and length scale $L = 70$~mm.  $\mu/\Rey$ is the intrinsic particle
  inertia, stripping out the variable component of fluid inertia}
\label{tab:properties}
\end{center}
\end{table}

Particle properties are given in Table~\ref{tab:properties}.  The
largest particle is the light capsule used for orbit visualization,
described below.  Experiments to measure the clustering instability
boundary and the clustering rate use the other four particles.  The
light capsules used in figure~\ref{fig:orbits} have $\mu \sim 1$, and
the particles used to produce the data in Figs.~\ref{fig:results} and
\ref{fig:rate} have $10^{-4} < \mu < 10^{-2}$.

\subsection{Visualization}

\begin{figure}
\centering
\begin{tabular}{c}
\includegraphics[width=0.9\columnwidth]{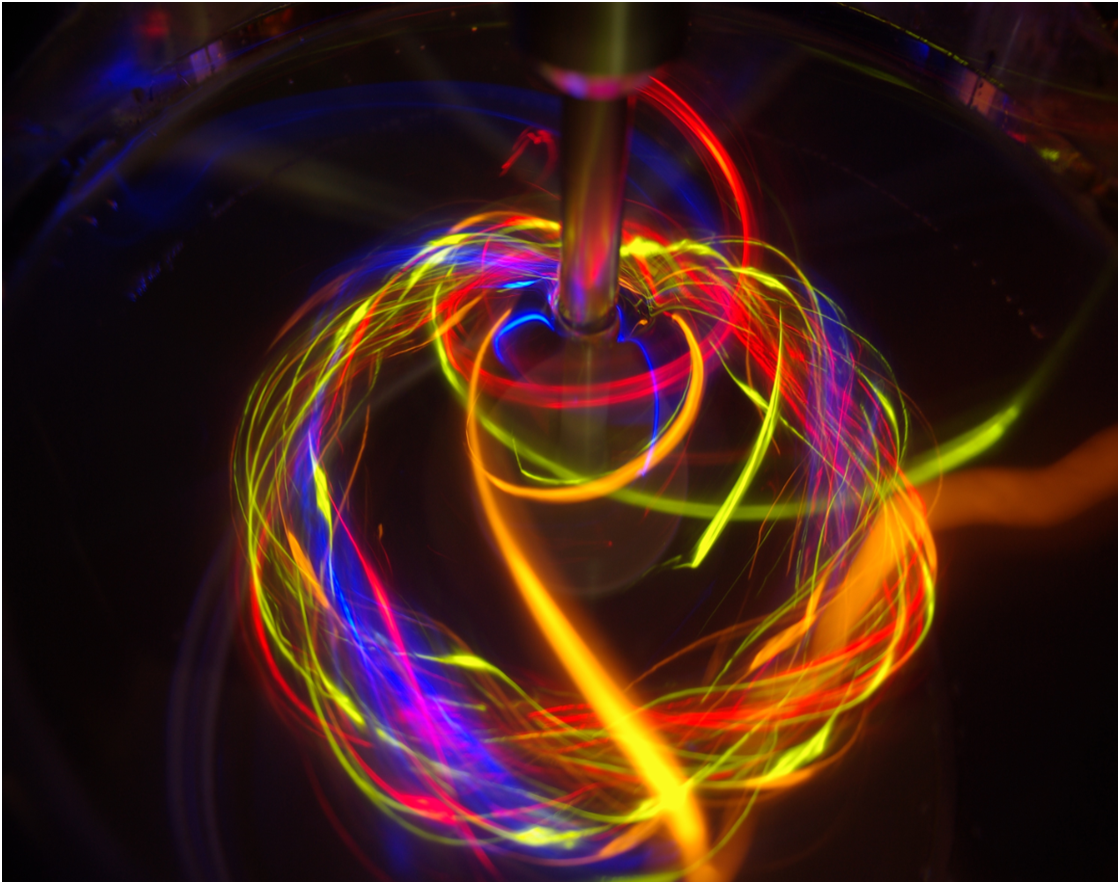} \\
(a) \\[1.5em]
\includegraphics[width=0.9\columnwidth]{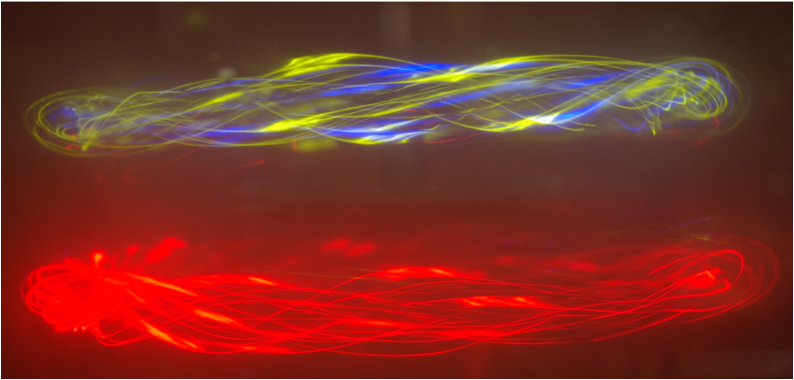} \\
(b)
\end{tabular}
\caption{Chaotic flow in a tank with impeller.  Orbits visualized with
  long exposure photographs using light capsules of various colors.
  (a) Capsules started near the fluid surface, showing spontaneous
  migration of particles into a KAM tube.  (b) Side view and close-up
  of impeller region; impeller not visible.  Above and below impeller
  are KAM tubes in which particles execute helical orbits.}
\label{fig:orbits}
\end{figure}

We visualize particle and fluid orbits in several ways.  A new
technique that we developed is particularly good at exposing the
structure of individual orbits and is also suitable for larger
containers.  This method places a light emitting diode inside a
transparent spherical capsule.  Figure~\ref{fig:orbits} shows long
exposure photographs after one or a few light capsules were placed in
the tank.  Fig~\ref{fig:orbits}a is an oblique view from above and
figure~\ref{fig:orbits}b is a side view.  Above and below the impeller
are the KAM tubes inside of which the light capsules trace out helical
orbits.  Outside the torus the pathlines of fluid particles are
chaotic and well-mixed.  The flow structure in the laminar flow tank
is a typical chaotic flow with coexisting KAM and chaotic regions.
This is the kinematic flow template that mediates the effects of
inertial forces.

Several groups have examined particle motion {\em inside} laminar flow
vortices
\cite{Rudman_AIChEJ_1998,Wereley_inertial_1999,Adetola_disks_2006}
and found that particles migrate toward the center of a tube or, if
cantori exist, particles may migrate into the cantori.
Figure~\ref{fig:orbits}a shows orbits three capsules initially placed
at the fluid surface that spontaneously and rapidly move through the
chaotic region into the tube; once in the tube the capsules move in
helical orbits passively with the fluid flow.  Below we quantify and
explain this type of particle localization using the four low inertia
particles in Table~\ref{tab:properties}.  To our knowledge, our
experiments are the first to show how inertial particles get into
laminar flow vortex tubes.

\section{Clustering Instability}
\label{sec:instability}

\begin{figure}
\centering
\begin{tabular}{cc}
\includegraphics[width=0.45\columnwidth]{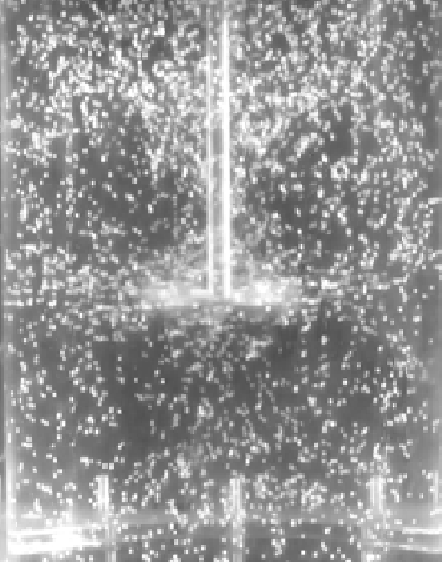} &
\includegraphics[width=0.45\columnwidth]{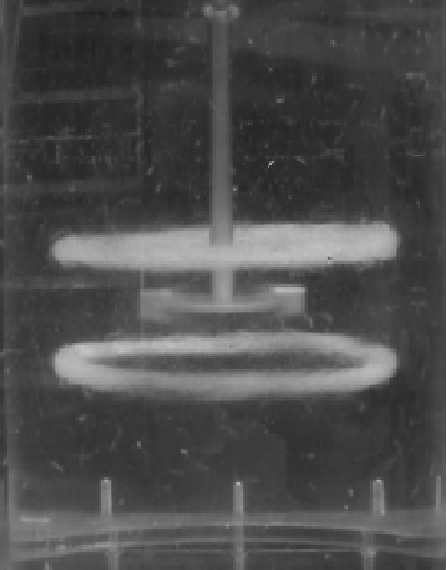} \\
before instability & after instability
\end{tabular}
\caption{(right) After instability particle cluster into KAM tubes.
  (left) Before instability particles move throughout the tank except
  in KAM tubes.}
\label{fig:results}
\end{figure}

Figure~\ref{fig:results} illustrates the clustering transition.  Flow
is established at a particular $\Rey$, then particles are poured into
the tank and swiftly disperse in the fluid everywhere {\em except}
into the KAM tubes (figure~\ref{fig:results}a): particles move
sufficiently passively to respect the material surface.  At a somewhat
higher $\Rey$ we find that the particles spontaneously cluster into
the tubes (figure~\ref{fig:results}b).  There are two necessary
conditions for particle clustering: (1) the existence of repellors to
convert particle trajectories from being nearly coincident with fluid
trajectories (passive behavior) into particle trajectories that are
nearly independent of fluid trajectories (non-passive behavior); and,
(2) separated flow regions, not too far away from repellors, into
which particles can settle and that prevent particles from
encountering repellors again.  As condition (2) is amply fulfilled by
the tubes of figure~\ref{fig:orbits}, we now specify condition (1) for
a fluid repellor.

Prior to the work of Cartwright, Piro and co-workers
\cite{Babiano_small_2000,2002_Cartwright_bailout}, it was commonly
assumed that particles with low enough inertia would behave passively,
i.e.\/ a particle released in a fluid with a slight velocity mismatch
compared to the fluid velocity (initially small $\Rey_p$) would adapt,
after a short transient, to the local fluid velocity.  If the right
side of (\ref{eq:MR_reduced}) is small, then the assumption appears
natural to take $D/Dt = d/dt$, in which case (\ref{eq:MR_reduced}) has
the trivial solution that an initial velocity mismatch decays
exponentially in a characteristic time of $\mu$.  However, this is not
always true.  Sapsis and Haller
\cite{2008_Sapsis_instabilities,2008_Haller_inertia,Sapsis_criterion_2010},
expanding on the earlier work
\cite{Babiano_small_2000,2002_Cartwright_bailout}, proved that the
fluid plane of (\ref{eq:MR_reduced}) is overall attracting, but also
that parts of the plane repel particles wherever
\begin{equation}
\label{eq:math_criterion}
\sigma_{min} \left[ \mathbf{S} + \mu^{-1} \mathbf{I} \right] < 0.
\end{equation}
$\mathbf{S} = \frac{1}{2}\left[\nabla \mathbf{u} + \left(\nabla
    \mathbf{u}\right)^T\right]$ is the rate of strain tensor,
$\mathbf{I}$ is the unit tensor and $\sigma_{min}\left[ \cdot \right]$
denotes the minimum eigenvalue of the bracketed tensor.  As our flow
is axisymmetric, the required eigenvalue can be determined from the
characteristic equation of the $2\times2$ matrix in
(\ref{eq:math_criterion}) as $\sigma_{min} = \mu^{-1} -
(-\det(\mathbf{S}))^{1/2}$.  From a general result of matrix algebra,
$-\det(\mathbf{S}) = \frac{1}{2} \mathrm{tr}(S^2) =
\frac{1}{2}\mathbf{S}:\mathbf{S}$, where we have used that
$\mathrm{tr}(\mathbf{S}) = 0$; the dyadic (double dot) product gives
the sum of the squared strain rates along the eigendirections of
$\mathbf{S}$.  The invariant dyadic product of the rate of strain
tensor is often used as the definition of the (local) total strain
rate \cite{Chong_general_1990,Speetjens_regime_2006}, and we define
the total strain rate as $\dot{\gamma} =
\frac{1}{2}(\mathbf{S}:\mathbf{S})^{1/2}$, which converts
(\ref{eq:math_criterion}) to
\begin{equation}
\label{eq:simple_criterion}
\mu \dot{\gamma} > 1,
\end{equation}
which we interpret as inertial stress.  As the strain along
streamlines is proportional to the radius of curvature of the
streamline, a particle (nearly) passively follows a fluid streamline
until the streamline curves too much, with (\ref{eq:simple_criterion})
quantifying how much strain is ``too much''.  As a particle flows
around the streamline curve, if the stress required to follow the
fluid streamline exceeds (\ref{eq:simple_criterion}), then inertial
perturbations amplify particle trajectories away from fluid
trajectories.

Strain rate varies with position in the fluid, and
(\ref{eq:simple_criterion}) is a local criterion.  We estimate strain
in the tank from the literature \cite{handbook}.  For Newtonian fluid
$\dot{\gamma} = k \Rey$, where $k$ is a constant specific to impeller
type.  Equation~\ref{eq:simple_criterion} then becomes
\begin{equation}
\label{eq:final_criterion}
\mu k > \Rey^{-1},
\end{equation}
which is no longer a local quantity.  Rather, it is an upper bound to
the global instability boundary of critical Reynolds number for a
particle with inertia $\mu$ to scatter from repellors.

\begin{figure}
\centering
\includegraphics[width=0.95\columnwidth]{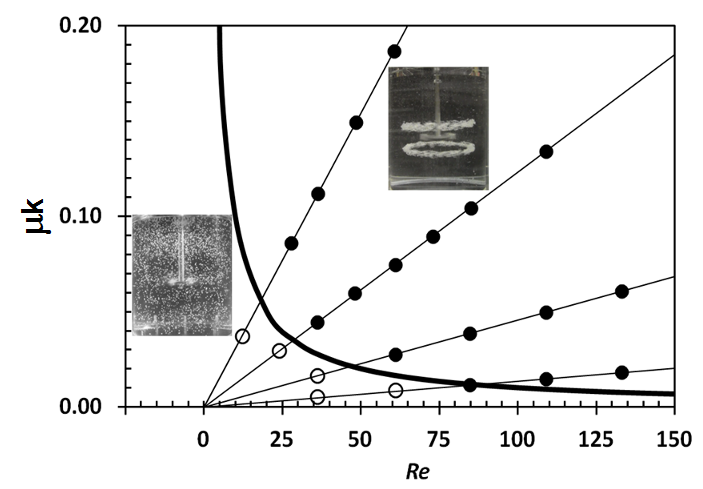}
\caption{Instability boundary theory and data.  As a function of $\Rey$
  each particle type goes along a diagonal line with slope
  proportional to inertia.  Thick line is
  (\ref{eq:final_criterion}).  Open circles, no clustering; filled
  circles, clustering.  For a Rushton impeller $k = 5$.}
\label{fig:compare}
\end{figure}

With the two necessary clustering conditions in hand, our experimental
procedure is to introduce particles into the established flow at low
$\Rey$, increase $\Rey$ incrementally and film activity in the tank.
Figure~\ref{fig:compare} plots data for four particles with $10^{-4} <
\mu < 10^{-2}$.  Open circles denote no clustering and solid symbols
clustering.  As $St$ contains a factor of $\Rey$, each particle plots
along a diagonal line with slope proportional to inertia.  The thick
solid line is (\ref{eq:final_criterion}).  To the right and above the
line, (\ref{eq:final_criterion}) predicts clustering, which is
confirmed by the data.  It is remarkable that the data agree so well
with (\ref{eq:final_criterion}), which has no adjustable parameters.

\section{Clustering Rate}
\label{sec:rate}

Once a particle encounters repellor(s), it acquires a non-zero
$\Rey_p$ and moves away from the fluid plane to move in the full
inertial phase space (figure~\ref{fig:skeleton}b).  As the augmented
dynamical system is non-conservative, the particle will reattract onto
the fluid subspace, i.e.\/ through drag the particle's velocity and
path will eventually approximate those of some fluid streamline until
a repellor is again encountered.  It is tempting to suppose that the
{\em clustering rate} of particles into tubes is given by the
divergence (\ref{eq:divergence}).  The clustering rate should be
proportional to the divergence; however, (\ref{eq:divergence}) is not
the correct divergence because a repelled particle has $\Rey_p > 1$
and the MR equation is no longer valid.  Prediction of the rate of
particle capture breaks down into two related questions: what is the
probability $P_c$ of a particle being captured into a tube; and, what
is the probability distribution $\cal P$ for the location in the fluid
where a particle returns to moving passively?


\begin{figure}[t]
\centering
\begin{tabular}{cc}
\multicolumn{2}{c}{\includegraphics[width=0.55\columnwidth]{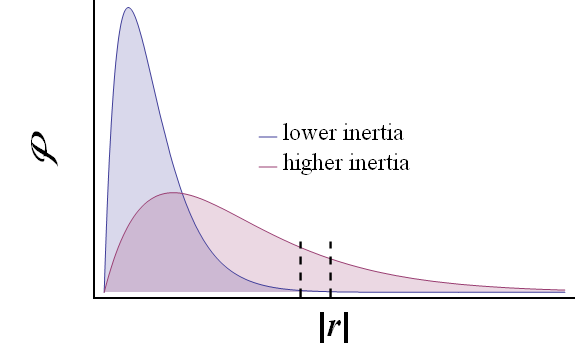}} \\
\multicolumn{2}{c}{(a)} \\[2em]
\raisebox{-0.5\height}{\includegraphics[width=0.477\columnwidth]{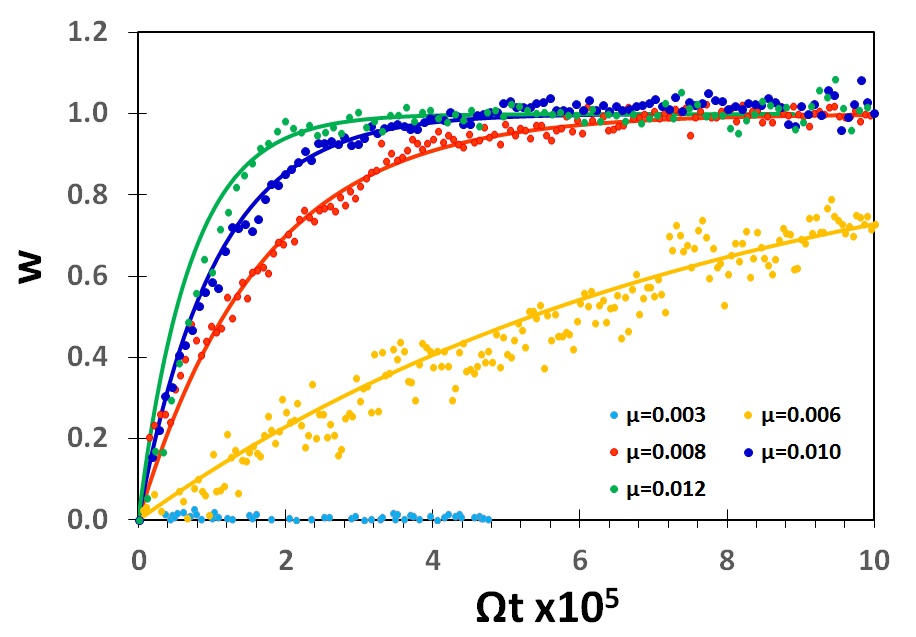}} &
\raisebox{-0.5\height}{\includegraphics[width=0.446\columnwidth]{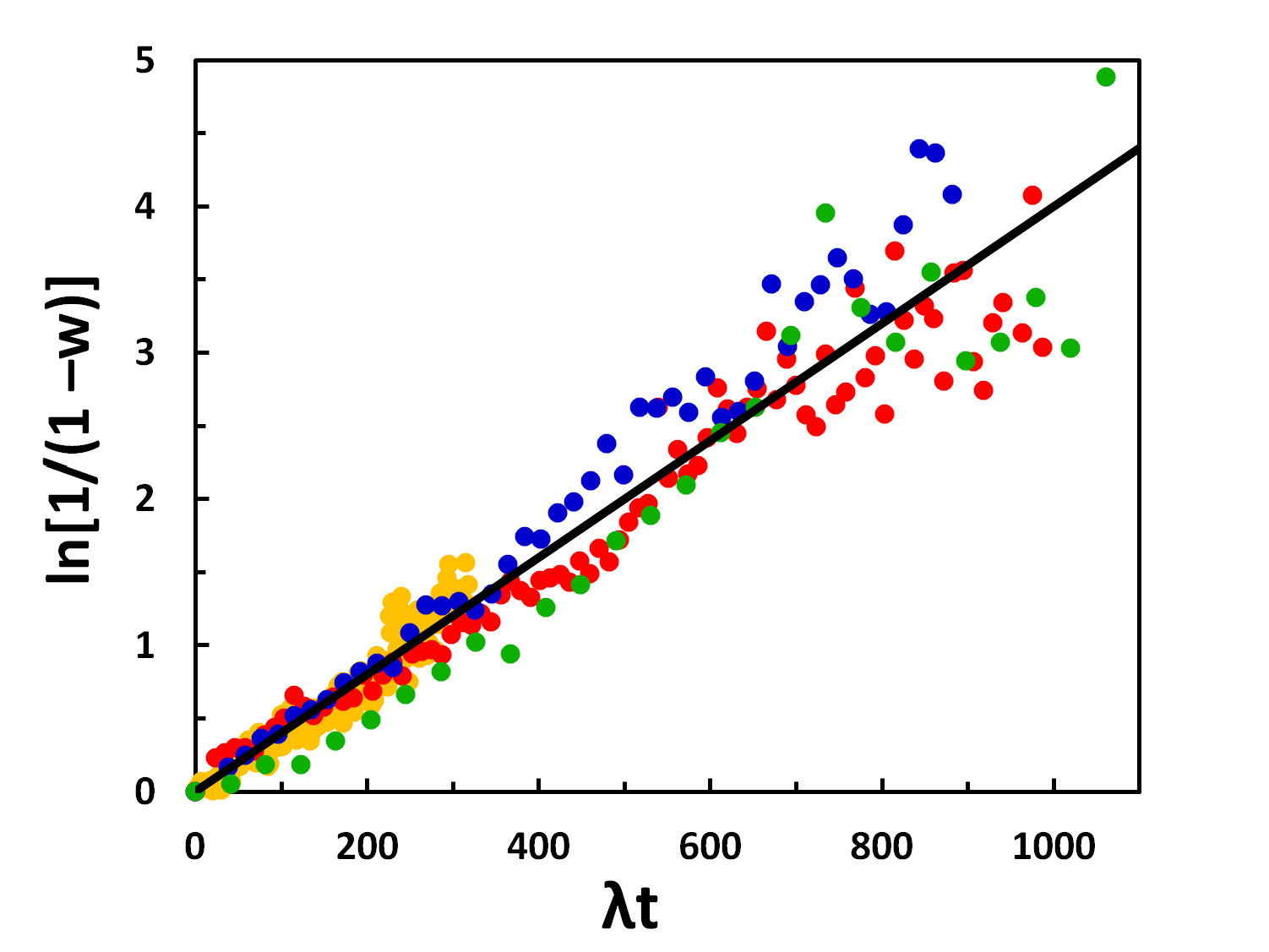}} \\
(b) & (c)
\end{tabular}
\caption{Clustering rate.  (a) Illustration of the probability density
  function ${\cal P}$ versus distance from a repellor.  Vertical
  dashed lines illustrate a roll.  (b) The increase of number of
  clustered particles as a function of time non-dimensionalized by the
  impeller rotation rate at the indicated inertia.  Dots are data and
  lines are exponential fits.  Cyan points are below the clustering
  threshold.  (c) Data in (b) collapse onto a universal curve when
  time is scaled by the rate $\lambda$ defined in the text.  Colors
  follow the legend in (b).}
\label{fig:rate}
\end{figure}

If $N_t$ is the number of particles that have gone into the tube, then
the increase of $N_t$ during a time interval $dt$ is $dN_t = P_c \:
dt$.  The capture probability per unit time is proportional to the
number of particles still outside the tube, $N_f$ (with the total
number of particles $N_0 = N_f + N_t$) and to the circulation rate of
the flow.  The capture probability is also proportional to the
convolution of $\cal P$ with the tube volume fraction $\cal V$, which
we have independently measured as a function of $\Rey$.  Even in the
absence of detailed knowledge of $\cal P$ we can still reach some
general conclusions.  Particles with less inertia will on average
reacquire a fluid streamline closer to the repellor.  With increasing
particle inertia, $\cal P$ will spread out more evenly over the fluid
domain.  Figure~\ref{fig:rate}a illustrates this idea, in which
$|\mathbf{r}|$ is distance from the repellor and the dashed lines
represent a separated region.  From this line of reasoning we expect
the clustering rate to increase with inertia.  Of course, this
illustration will be complicated by the possibility of more than one
repellor.  The capture probability per unit time $P_c \propto \Omega
N_f \int {\cal P} {\cal V}\; dv$, where the convolution integral is
taken over the fluid domain, and it follows that
\begin{equation}
\label{eq:rate_equation}
\frac{d N_t}{d \tau} \propto
          \left(\frac{1}{2}N_0 - N_t\right) \int {\cal P} {\cal V}\; dv,
\end{equation}
where $\tau = \Omega t$ and the $1/2$ is from measuring the capture
into a single tube whereas there are two tubes.  Defining $w =
N_t/(N_0/2)$, the solution to (\ref{eq:rate_equation}) is
\begin{equation}
\label{eq:rate_solution}
w = 1 - \exp(-\lambda t).
\end{equation}
$N_t$ is an exponential function of time that saturates at a rate
$\lambda = \Omega \int {\cal P} {\cal V}\; dv$.

We test this prediction by measuring the number of particles captured
by a tube $w$ as a function of time $t$.  We do this by measuring the
light transmitted through the region of a single tube and summing the
pixel intensity.  Figure~\ref{fig:rate}b shows that $w$ is indeed a
saturating exponential and that the clustering rate increases with
$\mu$.   We find that
$\lambda = C (\mu - \mu_c)^\beta {\cal V} \Omega$,
with $\mu_c$ the value of $\mu$ at the instability boundary in
figure~\ref{fig:compare}, collapses all data onto a universal curve
(figure~\ref{fig:rate}c) with $\beta = 1.79 \pm 0.03$ and $C = (4.01
\pm 0.04) \times 10^{-3}$.

\section{Conclusions}
\label{sec:conclusions}

We have outlined experiments and theory that pose some fresh answers
to the question, where do particles go in fluid flows?  As the
ultimate fate, location and arrangement of particles is a key
determinant of material form and function and of industrial process
efficiency, this is a question that arises in diverse areas.  Taking a
dynamical systems approach, we have shown that the kinematics of
inertial particles given by the MR equation has an augmented phase
space with a hyperplane corresponding to passive fluid motion that
both attracts and repels particles.  The interplay of attractors and
repellors can localize particles into small volumes of the fluid
domain.  Physically these involve respectively separated and
sufficiently straining flow regions.
Equation~\ref{eq:final_criterion} gives the critical $\Rey$ for when
particle paths become divorced from fluid streamlines which is the
onset of the clustering instability; the instability boundary depends
only on material properties and the characteristic strain rate.  MR
kinematics accurately predicts how particles starting on the fluid
manifold leave but cannot predict how particles reattract onto the
fluid manifold.  However, the clustering rate is well-characterized by
a scaling relation.  As the two necessary physical attributes are
readily found in chaotic flows, we anticipate being able to activate
this clustering instability in many flows.


Future work will take several approaches.  First measurement or
computation of the details of the velocity field would map out the
location(s) and spatial extent of repellors.  In the same vein direct
measurement of the probability distribution $\cal P$ would be a
difficult 3-dimensional experiment but could be done to understand the
physics determining the clustering rate.  Also, MR theory is
restricted to spherical particles; extension of theory to
non-spherical or deformable particles would be of great use.  Finally,
it has not escaped our notice that the clustering phenomena we have
exposed suggests a basis for separation technologies.

\bibliographystyle{plain}

\bibliography{Cluster}

\end{document}